\documentclass[prb,aps,showpacs,twocolumn,floatfix]{revtex4}
\usepackage{graphicx}
\usepackage{dcolumn}
\usepackage{bm}
\usepackage{epstopdf}

\begin{document}

\title{Effects of charged defects on the electronic and optical properties of self-assembled quantum dots}

\author{Ranber Singh and Gabriel Bester}
\affiliation{Max-Planck-Institut f\"{u}r Festk\"{o}rperforschung, Heisenbergstrasse 1, 70569 Stuttgart, Germany.}

\date{\today}

\begin{abstract}
We investigate the effects of point charge defects on the single particle electronic structure, emission energies, fine structure splitting and oscillator strengths 
of excitonic transitions in strained In$_{0.6}$Ga$_{0.4}$As/GaAs and strain-free GaAs/Al$_{0.3}$Ga$_{0.7}$As quantum dots.
We find that the charged defects significantly modify the single particle electronic structure and excitonic spectra in both strained and strain-free structures.
However, the excitonic fine structure splitting, polarization anisotropy and polarization direction in strained quantum dots remain nearly unaffected, 
while significant changes are observed for strain-free quantum dots.
\end{abstract}
\date{\today}
\maketitle
\section{introduction}

Point charge defects in semiconductors modify their properties, including their electronic structure, trapping and recombination rates for electrons and holes, and luminescence quenching \cite{seebauer06}. Point defects in semiconductors have been studied extensively with the aim to use their properties for defect engineering.
Understanding the properties of point defects is important in studying the electrical and optical properties of semiconductors. 

To date, little is known about the influence of point defects on the electronic structure and optical properties of self-assembled  semiconductor quantum dots (QDs) grown epitaxially by the Stranski-Krastanow growth mode.
These type of QDs provide a quantum system which can be engineered to have a wide range of desired properties \cite{michler00,cortez02,michalet05,gerardot08,patel10} and  
provide a rather unique opportunity to investigate various effects known from atomic physics, in a solid state environment \cite{kroner08,santori09}.
However, their electronic structure and optical properties can be altered by the presence of defects inside the QDs 
or in the surrounding barrier \cite{chi04}.
The defects responsible for the electron (hole) trapping in III-V semiconductor heterostructures are the positively (negatively) charged acceptor (donor)-like defects such as vacancies, interstitial or substitutional defects \cite{deppe88}.
 These defects in strained In(Ga)As/GaAs and strain-free GaAs/Al(Ga)As QDs may develop during the growth process \cite{meng07,belyaev00}. The negatively charged donor-like substitutional Si defect in GaAs QDs are predicted to be stable \cite{jingbo05}.
A point charge may also get trapped at a defect site during an optical or electrical measurement. Charged defects inside a QD or in the surrounding region may change the depth and symmetry of the confining potential quite significantly  
affecting the electronic and optical properties of the QD.
The influence of these defects on the optical properties have been shown only in few cases, namely for self-assembled Ge/Si \cite{nguyen06}, CdSe/ZnSe \cite{semenova02}, and InGaAs/AlGaAs \cite{kamada08} QDs.
For colloidal QDs, on the other hand, the phenomena has been extensively investigated.
The so-called blinking, is often attributed to the trapping of carriers into defect states on the surface or the surrounding matrix \cite{efros97,klimov00} and represents a severe limitation for many potential applications\cite{mahler08}.  
For self-assembled QDs, we may speculate that charged defects may be responsible for the large spread of the excitonic fine structure splitting (FSS), polarization direction and polarization anisotropy in strained In(Ga)As and strain-free GaAs/Al(Ga)As QDs \cite{seidl08,plumhof10,seguin05,lin11}.

We investigate the effects  of point charge defects on the electronic structure, the FSS, emission energies and oscillator strengths of optical transitions in strained In$_{0.6}$Ga$_{0.4}$As/GaAs and strain-free GaAs/Al$_{0.3}$Ga$_{0.7}$As QDs. We use an atomistic empirical pseudopotential description and configuration interaction. The charged defects are modeled at the level of the hydrogenic effective mass description (neglecting central cell effects). We find that the charged defects significantly modify the electronic structure and excitonic spectra in both strained and strain-free QDs. 
However, the FSS, polarization anisotropy and polarization direction in strained QDs remain nearly unaffected, while in strain-free QDs there are significant changes in these quantities.  

\section{Computational details}

\begin{figure}[h]
\begin{centering}
\includegraphics[width=\linewidth]{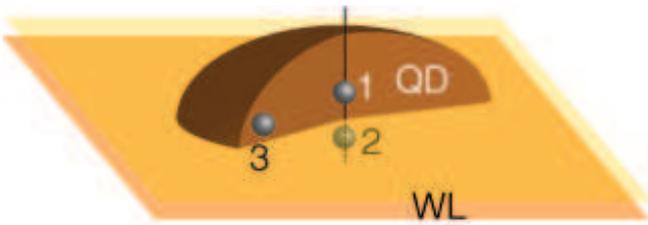}
\caption{Position of different point charges; 1) near the QD center, 2) in the WL underneath the base of the QD and 3) near the QD edge along the [1$\bar{1}$0] crystal direction.} \label{fig:Charged_Geometry}
\end{centering}
\end{figure}

We consider impurity defects \cite{montenegro92} given as, $V=\frac{\pm1}{\epsilon |{\bf r}-{\bf r}_0|}$, following hydrogenic effective mass theory and 
ignoring the central cell effects \cite{pantelides78,martins02}, where $\epsilon $ is the dielectric constant of the material composing the QD and ${\bf r}_0$ is the position of the impurity. 
The positively ($\oplus$) and negatively ($\ominus$) charged point defects are placed at different locations shown schematically in Fig.\ref{fig:Charged_Geometry}: 1) inside the QD near its center, 2) 
in the wetting layer (WL) underneath the base of the QD, 3) inside the QD near the interface with the barrier along the [1$\bar{1}$0] direction. 
We used a lens-shaped QD with cylindrical base of  
25.2 nm diameter and 3.5 nm height. We relax the atomic positions within the simulation cell to the minimum strain energy using the valence force field method \cite{keating66,williamson00}. 
The single particle orbitals and energies of the QD are calculated by using the atomistic empirical pseudopotential approach 
\cite{wang99b,williamson00}, taking strain, band coupling, coupling between different parts of the Brillouin zone and spin-orbit 
coupling into account, retaining the atomistically resolved structure. The Coulomb and exchange integrals are calculated from the atomic 
wave functions as shown in Ref.\onlinecite{bester03} and the correlated excitonic states are calculated by the configuration interaction 
(CI) approach \cite{franceschetti99}. For the CI calculations we use all possible determinants constructed from the twelve lowest energy 
electron and twelve lowest hole states (spin included), thus accounting for correlations. 

\section{Results and discussion} 

\subsection{Single particle states}

\begin{figure}[t]
\begin{centering}
\includegraphics[width=\linewidth]{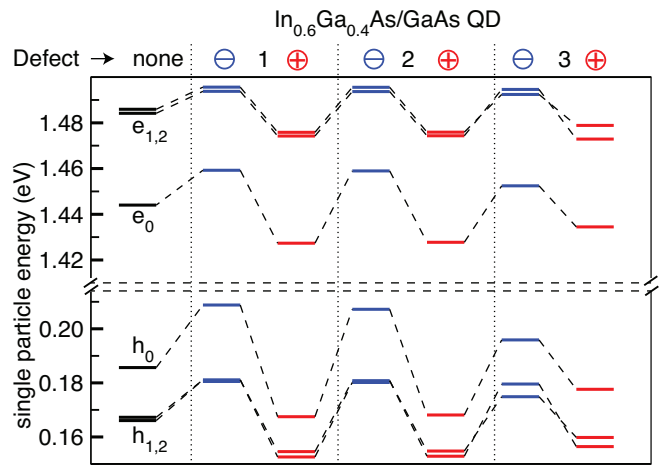}
\caption{Single particle energies (relative to the valence band maximum of bulk GaAs) of electron ($e_0$, $e_{1,2}$) and hole ($h_0$, $h_{1,2}$) states for different point charge defects
in In$_{0.6}$Ga$_{0.4}$As/GaAs QDs. The position of the defects (1, 2, 3) are defined in Fig.\ref{fig:Charged_Geometry}.} \label{fig:SP_InGaAs}
\end{centering}
\end{figure}

\begin{figure}[ht]
\begin{centering}
\includegraphics[width=\linewidth]{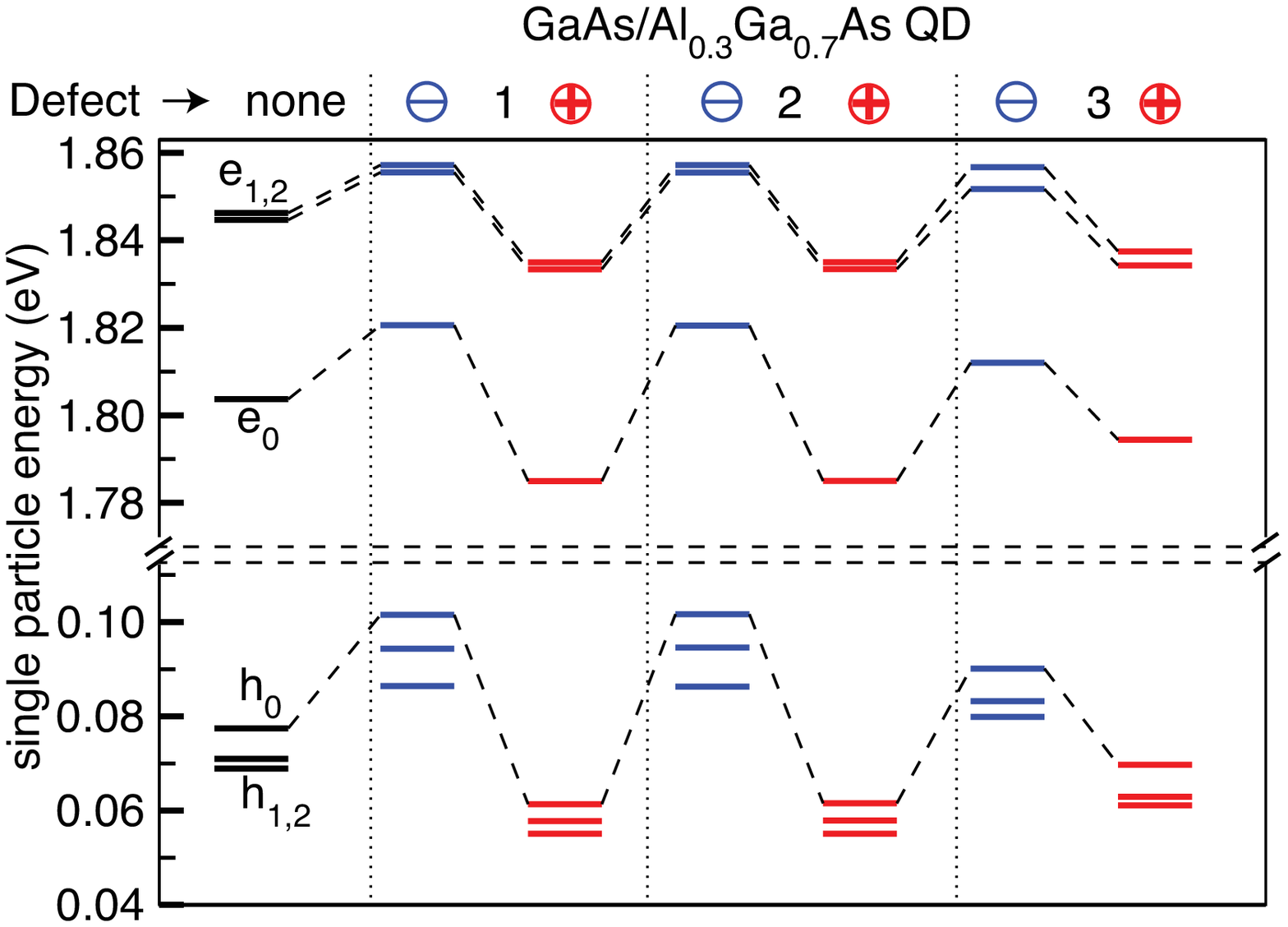}
\caption{Single particle energies (relative to the valence band maximum of bulk GaAs) of electron ($e_0$, $e_{1,2}$) and hole ($h_0$, $h_{1,2}$) states for different point charge defects
in GaAs/Al$_{0.3}$Ga$_{0.7}$As QDs. The position of the defects (1, 2, 3) are defined in Fig.\ref{fig:Charged_Geometry}.}  \label{fig:SP_GaAs}
\end{centering}
\end{figure}

The single particle energies of electron ($e_{0,1,2}$) and hole ($h_{0,1,2}$) states are given in Figs. \ref{fig:SP_InGaAs} and \ref{fig:SP_GaAs}
for the different defect configurations in In$_{0.6}$Ga$_{0.4}$As/GaAs and GaAs/Al$_{0.3}$Ga$_{0.7}$As QDs.
We observe significant changes in the electronic structure of both strained In$_{0.6}$Ga$_{0.4}$As 
and strain-free GaAs QDs in response to the charged defects. The energetic splitting between $S$ and $P$ states is significantly modified and the energies of electron 
and hole states blue-shift (red-shift) for negatively (positively) charged defects. This is due to the fact that the negative defects repel (attract) an electron (hole) 
and accordingly for the positive defects. The energetic splittings between the electron $P$ states ($e_{1,2}$) and holes ($h_{1,2}$) increase when the defects are placed along the [1$\bar{1}$0] crystallographic direction because it increases the 
inequivalence between the [110] and [1$\bar{1}$0] directions. In strain-free GaAs QDs there is a larger splitting of the $P$ hole states for the negative defects placed 
at positions 1 and 2. This is due to the increase of light-hole character in response to the delocalization of the wave functions in z-direction as will be demonstrated next. 
The decomposition of the S hole state into light-hole (LH) and heavy-hole (HH) components is 
given in Table \ref{tab:table1}. The negative point defects $\ominus$ at positions 1 and 2 lead to an increase of the LH-component, while the positive point defects  
$\oplus$ at the same positions induce a reduction of the LH-component. This can be understood from the reduced and increased Z-component of the hole wave functions. 
The HH is favored when the spread of the wave function in z-direction is small. The introduction of the positive point charge $\oplus$ in the center of the dot leads 
to a delocalization in-plane and a reduction of the z-character of the wave functions, and hence of its LH-component. The negative point charge $\ominus$ attracts 
the wave function that becomes more spherical with an increase of its z-character, and hence increase the LH-component

\begin{table}[ht]
\caption{Decomposition of $S$ hole states into LH and HH components and Coulomb integrals of hole-hole (J$_{hh}$), hole-electron (J$_{he}$) and electron-electron (J$_{ee}$) interactions for the $S$ states.
The sign `+' (`-') indicate positively (negatively) charged defect. The position of defects (1, 2, 3) are defined in Fig.\ref{fig:Charged_Geometry}.}
\centering
\begin{ruledtabular}
\begin{tabular}{l|l|ccccc}
&&\multicolumn{2}{c}{$S$ hole state}&\multicolumn{3}{c}{Coulomb integrals}\\
QD&defect &HH&LH& J$_{hh}$&J$_{he}$&J$_{ee}$\\\hline
      &none         & 89.48 & 09.33 & 25.97 &  24.86 &  23.94 \\
      &1$^{'-'}$ & 84.50 & 13.90 & 35.37 &  26.11 &  22.26 \\
      &2$^{'-'}$ & 84.16 & 14.21 & 35.61 &  26.15 &  22.28 \\
GaAs  &3$^{'-'}$ & 84.77 & 13.76 & 29.50 &  20.54 &  24.01 \\
      &1$^{'+'}$ & 90.73 & 08.10 & 20.65 &  22.15 &  25.98 \\
      &2$^{'+'}$ & 90.22 & 08.61 & 20.78 &  22.26 &  25.98 \\
      &3$^{'+'}$ & 88.14 & 10.65 & 26.17 &  23.82 &  23.89 \\\hline
      &none         & 94.00 & 02.46 & 24.68 &  20.95 &  18.61 \\
      &1$^{'-'}$ & 92.70 & 03.31 & 29.91 &  21.03 &  17.35 \\
      &2$^{'-'}$ & 92.87 & 03.21 & 29.34 &  21.00 &  17.40 \\
In$_{0.6}$Ga$_{0.4}$As&3$^{'-'}$&93.90  &02.52  &  24.52 & 19.95 & 18.64 \\
      &1$^{'+'}$ & 94.34 & 02.25 & 21.13 &  20.12 &  20.05 \\
      &2$^{'+'}$ & 94.36 & 02.23 & 21.26 &  20.29 &  19.97 \\
      &3$^{'+'}$ & 93.86 & 02.56 & 25.10 &  20.30 &  18.67 \\
\end{tabular}
\end{ruledtabular}
\label{tab:table1}
\end{table}

\begin{figure}[h]
\begin{centering}
\includegraphics[width=\linewidth]{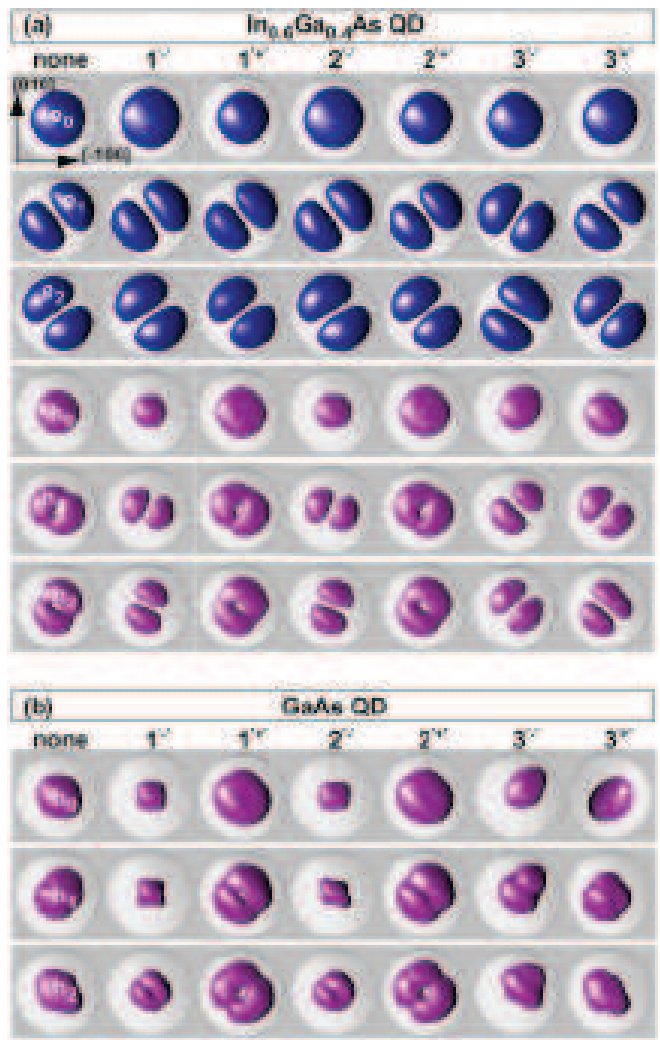}
\caption{Square of the wave functions for different point charge defects for In$_{0.6}$Ga$_{0.4}$As/GaAs (a) and GaAs/Al$_{0.3}$Ga$_{0.7}$As (b) QDs. 
The isosurfaces enclose 75\% of the probability densities. The position of the defects (1, 2, 3) are defined in Fig.\ref{fig:Charged_Geometry}.}  \label{fig:SP_Wfn}
\end{centering}
\end{figure}

The isosurfaces enclosing 75\% of the probability density of electron and hole states in strained In$_{0.6}$Ga$_{0.4}$As/GaAs and strain-free GaAs QDs
with different point charge configurations are shown in Fig. \ref{fig:SP_Wfn}. The defect placed along the [1$\bar{1}$0] crystal direction (position 3) breaks the symmetry,
which is clearly evident from the spacial distribution of the hole states. The negative defect placed along the [1$\bar{1}$0] crystal direction changes the orientation of the $P$ states
as compared to the case of the defect-free QD. The negative (positive) defect at position 3 favors hole (electron) $P$ state along the [1$\bar{1}$0] crystal direction
and the effect is so large as to reverse the energetic order of the states. Similar effects are observed for electrons in strain-free GaAs/Al$_{0.3}$Ga$_{0.7}$As QDs (not shown).
The spatial extent of the hole states in the GaAs/Al$_{0.3}$Ga$_{0.7}$As QD deviates significantly from the nearly single-band case of the InGaAs/GaAs QD. 
This is a consequence of the strong band mixing observed in the GaAs QD and will be further discussed in section 
\ref{sec:excitonic_spectra}. We observe that the negative defects lead to 
a contraction of the wave functions, while the positive point charges close to the QD center (position 1 and 2) repel the wavefunctions. 

The Coulomb integrals for hole-hole (J$_{hh}$), hole-electron (J$_{he}$) and electron-electron (J$_{ee}$) interactions between the $S$ states are given in Table \ref{tab:table1}.
The increase (decrease) in the magnitude of the Coulomb integrals J$_{hh}$ and J$_{ee}$ indicates an increase (decrease) in the localization of the hole and electron states, respectively.
For the GaAs QD the presence of a  negative charge increases the $J_{hh}$ integrals by a large value of about 10 meV. The $J_{ee}$ integrals are thereby reduced by only around 1 meV. 
This is related to the rather different effective masses of the electron and the hole in this material. The hole state has an increased ability to localize in the additional Coulomb 
potential created by the charged defect, compared to the electron. The changes observed in the $J_{eh}$  Coulomb integrals follow qualitatively the  changes observed for the holes, i.e., 
for a negative defect, the magnitude of integrals increase. The same trends are observed for the InGaAs QD, but with a generally smaller magnitude.
\begin{table}[h]
\caption{Emission energy, FSS, polarization anisotropy and polarization direction with respect to the [110] crystal direction of neutral exciton.
The sign `+' (`-') indicate positively (negatively) charged defect. The position of the defects (1, 2, 3) are defined in Fig.\ref{fig:Charged_Geometry}. }
\centering
\begin{ruledtabular}
\begin{tabular}{l|l|cccc}
QD&defect & energy& FSS     & polarization	& polarization    \\
      &   &  (eV) &($\mu$eV)& direction ($^o$)  & (\%)  	  \\\hline
      &  none      &  1.700   &    12.2   & 89.8 &  07.5	       \\			   
      &1$^{'-'}$&  1.692   &    04.4   & 89.5 &  04.5	       \\			   
      &2$^{'-'}$&  1.692   &    11.9   & 00.1 &  03.5	       \\			   
GaAs  &3$^{'-'}$&  1.699   &    13.9   & 00.1 &  06.5	       \\			   
      &1$^{'+'}$&  1.700   &    17.1   & 89.9 &  11.9 	       \\			   
      &2$^{'+'}$&  1.700   &    22.7   & 89.9 &  14.9 	       \\			   
      &3$^{'+'}$&  1.699   &    17.9   & 90.0 &  12.2 	       \\\hline 		   
      & none       &  1.237   &     4.2  & 2.4  &  0.2      \\ 			
      &1$^{'-'}$&  1.229   &     3.5  & 3.5  &  0.1      \\ 			
      &$2^{'-'}$&  1.230   &     3.7  & 3.5  &  0.1      \\   
In$_{0.6}$Ga$_{0.4}$As&3$^{'-'}$&1.236&3.8&0.5 &1.2      \\
      &1$^{'+'}$&  1.239   &     5.0  & 2.2  &  0.1      \\
      &2$^{'+'}$&  1.238   &     4.7  & 1.8  &  0.1      \\
      &3$^{'+'}$&  1.236   &     4.3  & 2.4  &  0.1      \\
\end{tabular} 
\end{ruledtabular}
\label{tab:table2}
\end{table}

\subsection{Exciton fine structure splitting and polarization}
The emission energy, FSS, polarization direction and polarization anisotropy of neutral excitons in In$_{0.6}$Ga$_{0.4}$As/GaAs
and GaAs/Al$_{0.3}$Ga$_{0.7}$As QDs is tabulated in Table \ref{tab:table2}. The polarization anisotropy is defined as
(I$_{1}$-I$_{2}$)/(I$_{1}$+I$_{2}$), where I$_{1}$ and I$_{2}$ are, respectively, the intensities along major and minor axis of the elliptically 
polarized total emission of two neutral bright exciton transitions.
The FSS, polarization anisotropy and polarization direction remain almost unchanged in the presence of charged defects in In$_{0.6}$Ga$_{0.4}$As/GaAs QDs.
The polarization direction of the lower exciton is preferably along the [110] direction with small fluctuations in response to the defects. In pure InAs QDs 
the polarization direction of the lower exciton is along the [110] direction, which remains unaffected in the presence of different defect configurations (not shown).
However, there are changes in these quantities in strain-free GaAs/Al$_{0.3}$Ga$_{0.7}$As QDs. 
Polarization anisotropy decrease (increase) for the negative (positive) defects. The envelope functions of electron (hole) states are squeezed (expanded) 
for the positive defects. The contrary applies to the presence of negative defects (see Fig.\ref{fig:SP_Wfn}). 

The FSS is generally increased in case of positive defects while it is decreased in case of negative defects. This trend follows the observation of the electron-hole direct Coulomb 
interaction $J_{eh}$ in Table \ref{tab:table2}, where we argued that it is related to the  balance between the increase/decrease of electron and hole wave function localization. More explicitly, for negative 
defects the hole wave function becomes more localized while the electron wave function more delocalized. The effect on the hole overtakes the effect on the electron, and the Coulomb 
integral has a larger magnitude, compared to the defect-free case. This overall trend is rather intuitive, while the detailed quantitative results are more complicated: For negative defects in the GaAs QD,  
we obtain a strong reduction of the FSS when going from the defect free case to a defect at position 1 (12.2 to 4.4 $\mu$eV). Moving the defect to position 2 increases the magnitude 
of the FSS (11.2 $\mu$eV) but with reversed polarization, so the exciton lines have interchanged (anticrossed \cite{singh10}). We conclude, that the position of the defect along the symmetry axis 
(through the center of the dot, i.e., the z-axis) and the vertical electric field it creates, has a  significant influence on the FSS and is even able to ``reverse" it. Strong 
effects on the vertical electric fields on the FSS have been observed previously \cite{bennett10}.
The non-zero polarization anisotropy in GaAs QDs is due to large LH-HH mixing. We also notice that the polarization direction in GaAs QDs with 
cylindrical symmetry is not always along the [110] crystallographic axis. This is due to the atomistic random nature of the barrier composition and hence the interface of the QD.

\begin{figure}
\begin{centering}
\includegraphics[width=0.9\linewidth]{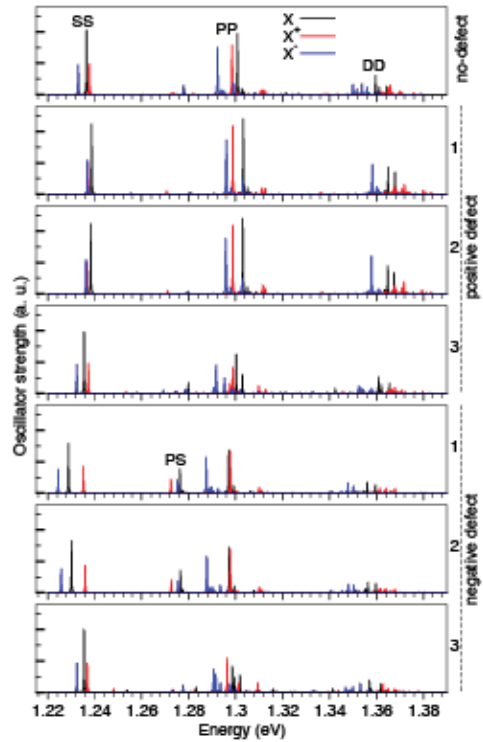}
\caption{Absorption spectra of neutral excitons (X) and charged excitons (X$^{+}$, X$^{-}$) in In$_{0.6}$Ga$_{0.4}$As/GaAs QDs.
The positions of the defects are defined (1, 2, 3) in Fig.\ref{fig:Charged_Geometry}. } \label{fig:exciton_InGaAs}
\end{centering}
\end{figure}

\begin{figure}
\begin{centering}
\includegraphics[width=0.9\linewidth]{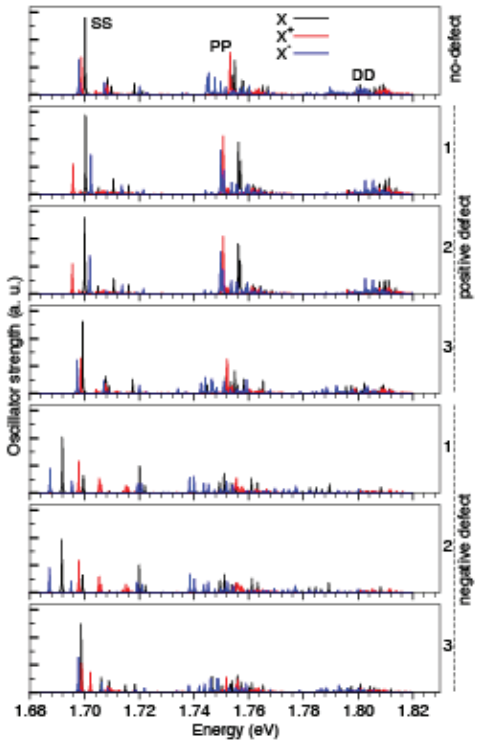}
\caption{Absorption spectra of neutral excitons (X) and charged excitons (X$^{+}$, X$^{-}$) in GaAs/Al$_{0.3}$Ga$_{0.7}$As QDs.
The positions of the defects are defined (1, 2, 3) in Fig.\ref{fig:Charged_Geometry}. The labels SS, PP and DD refer to the transitions from the electron S, P, D states, respectively, 
to the hole states h$_{0,1,2,3,4,5}$. Since the hole states have a heavily mixed orbital character (see Table \ref{tab:table3}), transitions to all of them are possible, 
leading to a high multiplicity of peaks.} \label{fig:exciton_GaAs}
\end{centering}
\end{figure}

\begin{table}
\caption{Decomposition of the atomistic hole wave functions in a GaAs QD into bulk  $\Gamma$ point Bloch functions times envelope functions: $\Psi_i({\bf r}) = \sum_n \phi_n({\bf r})
\psi_n({\bf r}) $ where $\phi_n$ is the Bloch function of band $n$ and $\psi_n$ is the corresponding
envelope function. The decomposition factors tabulated here are the {\it norm} of the envelope
functions for the three dominant valence bands $x$, $y$ and $z$. These bands are labelled by their axial angular momentum values $j = \frac{3}{2}$ for
$\frac{(x+iy)}{\sqrt{2}}\uparrow, \frac{(x-iy)}{\sqrt{2}}\downarrow$; $j=\frac{1}{2}$ for $\frac{(x-iy)}{\sqrt{2}}\downarrow, \frac{(x+iy)}{\sqrt{2}}\uparrow $ and 
$j=\frac{1}{2}^\prime$ for $z\uparrow, z\downarrow$. Each envelope function contains a combination of angular momenta (S, P, D), given in percent.}
\centering
\begin{ruledtabular}
\begin{tabular}{c|rrr|rrr|rrr}
      &   \multicolumn{9}{c}{norm of the envelope function} \\
      &    \multicolumn{3}{c}{$j=\frac{3}{2}$} & \multicolumn{3}{c}{$j=\frac{1}{2}$} & \multicolumn{3}{c}{$j=\frac{1}{2}^\prime$} \\
       label&  S   &  P   &  D   & S  & P & D & S & P  & D   \\\hline
 h$_0$      &  {\bf 83.1} &  2.3 &  0.1 &  0.3 & 0.6 & 0.6 & 0.5 &  2.0 &  1.1 \\ 
 h$_1$      &  3.6 & {\bf 40.2} & {\bf 6.4} &  4.5 & {\bf 7.4} & 0.4 & {\bf 8.2} & {\bf 13.6} &  0.7 \\ 
 h$_2$      &  3.5 & {\bf 15.0} & {\bf 15.6} & {\bf 11.9} & 4.4 & 0.2 &{\bf 22.4} &  {\bf 8.1} &  0.4 \\ 
 h$_3$      &  1.5 & {\bf 52.5} &  3.0 &  0.2 & 4.2 & 1.1 & 0.4 & {\bf 7.2} &  2.9 \\ 
 h$_4$      &  {\bf 34.0} & {\bf  5.0} &  2.6 &  0.5 & 1.6 & {\bf 9.9} & 0.7 &  3.1 & {\bf 18.4} \\ 
 h$_5$      &  {\bf 7.3} & {\bf 15.9} & {\bf 17.8} &  2.2 & {\bf 5.2} & 2.1 & 3.4 & {\bf 10.8} & 3.3 \\ 
\end{tabular}
\end{ruledtabular}
\label{tab:table3}
\end{table}

\subsection{Excitonic spectra}
\label{sec:excitonic_spectra}

The absorption spectra of neutral and charged excitons in InGaAs and GaAs QDs with different point charge defect configurations are given in Fig.\ref{fig:exciton_InGaAs} and \ref{fig:exciton_GaAs}.
The SS, PP and DD labels characterize the different orbital characters of the electron and hole states involved in the optical transition. 
The actual orbital S, P, D character of the hole wave functions for the GaAs QD is given in Table \ref{tab:table3}.
It becomes clear that strong deviations from orbitally pure states exist in GaAs. The states can hardly be classified as ``S", ``P", ``D" as 
they have a mostly mixed character. This is a direct consequence of the fact that the GaAs hole states have significant LH contributions due 
to the lack of strain in the structure (the LH band is closer to the HH band than in strained InAs dots). The hole states in GaAs QDs are even 
less ``single-band"-like than in strained InGaAs QDs. The consequence of this two-band character of the hole states in GaAs is not obvious from 
the single-particle energies (see Fig.\ref{fig:SP_GaAs}) but very clear from the absorption spectra in Fig.\ref{fig:exciton_GaAs}. 
The single-band model for holes, that leads to the popular simplified optical selection rules -- only SS, PP and DD transitions are allowed, leads to qualitatively 
wrong results. In GaAs, the mixed transitions ``SP", ``SD" where the hole is in a nominally P and D state are bright. Simply as a consequence, that these states have 
a significant S character, as given in Table \ref{tab:table3}.

The introduction of charged defects generally reduces the symmetry and introduces more LH-character in the hole states and more angular momentum mixing. As a consequence 
some transitions that were dark following the ``single-band optical selection rules" are now bright. This can be seen especially for the InGaAs QD in Fig.\ref{fig:exciton_InGaAs} 
that have relatively pure angular momentum hole states. In the defect free case, mainly the SS, PP and DD transitions are bright. The introduction of negative point charges 
leads to the appearance of nominally forbidden PS transitions. In the case of the GaAs QD the structure without defect has already a high level of angular momentum mixing 
(Table \ref{tab:table3}) that the effect is quantitative which leads to the appearance of new peaks in the spectrum. We note that the oscillator strength for the SS 
transition is only slightly reduced in the case of the negative defects placed close to the center of the QDs (position 1 and 2), for other cases it remains nearly unchanged. 

In summary, we investigate different point defect configurations in strained In$_{0.6}$Ga$_{0.4}$As/GaAs and strain-free GaAs/Al$_{0.3}$Ga$_{0.7}$As QDs. 
Depending upon the relative position of the charged defects, we obtain significant modifications in the single particle electronic structure and excitonic 
spectra in both strained and strain-free QDs. However, the FSS, polarization anisotropy and polarization direction in strained QDs remain nearly unaffected, 
while in strain-free QDs these quantities change significantly. We furthermore highlight that the hole states in GaAs QDs deviate significantly from a 
single-band object. We find LH components of up to 20\% and nominally ``P" states can have up to 30\% orbital S-character. This leads to a complex 
absorption picture with many peaks in GaAs QDs, compared to the case of strained InGaAs QDs.   

\begin{acknowledgements}
We would like to acknowledge financial support by the BMBF (QuaHL-Rep, Contract
No. 01BQ1034).
\end{acknowledgements}

\end{document}